\pdfoutput=1
\documentclass[aps,pra,reprint,showpacs]{revtex4-1}
\usepackage{graphicx}
\usepackage{amsmath}
\usepackage{amssymb}
\usepackage{amsfonts}
\usepackage{latexsym}
\usepackage{rotating}
\usepackage{color}
\usepackage{latexsym}
\usepackage{epsfig}
\usepackage{natbib}
\usepackage{bm}
\usepackage{url}

\usepackage{color} 

\usepackage{hyperref}
\hypersetup{
colorlinks=true,final=true,
        linkcolor=blue,
        citecolor=blue,
        filecolor=blue,
        urlcolor=blue,
}

\begin{document}

\title{Dissipative solitons in self-defocussing 
nonlinear media : The curious case of zero-nonlinearity point}
\author{Ambaresh Sahoo}
\email{ambaresh@phy.iitkgp.ernet.in}
\author{Samudra Roy}
\email{samudra.roy@phy.iitkgp.ernet.in}
\affiliation{Department of Physics, Indian Institute of Technology Kharagpur, West Bengal 721302, India}


\begin{abstract}

We theoretically model a dissipative system which exhibits self-defocussing nonlinearity and numerically study the dynamics of optical dissipative solitons (DSs) whose evolution is governed by a complex Ginzburg-Landau equation (GLE). We show that the formation of DSs is not restricted in the domain exhibiting positive nonlinearity. Stable DSs are excited even in the regime where the nonlinearity is negative. Based on the numeric sign of dispersion and nonlinear coefficient, we classify the operational regime into four discrete domains and study the formation of DSs in those regimes. We design a realistic waveguide that exhibits strong frequency dependent nonlinearity which changes its sign across a certain frequency called zero-nonlinearity point (ZNP). We adopt a variational technique to theoretically study the overall dynamics of DSs under various perturbations by choosing Pereira-Stenflo type soliton as our ansatz since it is the natural solution of the unperturbed GLE. An extensive numerical study reveals that the ZNP plays a dominant role on the pulse dynamics and depending on its relative location with respect to input frequency, it can either suppress or enhance Raman induced frequency down-shifting. This is further supported by the variational method which quantitatively determines the location of the Raman frequency as a function of the ZNP. The dispersive radiation generated due to third-order dispersion changes drastically with the location of the ZNP. We analytically derive a phase matching equation that predicts the location of radiation frequency in presence of the ZNP.

\end{abstract}

\maketitle

\section{Introduction}
\label{sec:intro}
Dissipative solitons (DSs) are stable localized structures that can be existed in a wide range of open nonlinear systems in nature \cite{Akhmediev}. Unlike the Schr\"{o}dinger Kerr solitons which is generated inside a conservative medium as a result of the balance between the group-velocity dispersion (GVD) and nonlinearity \cite{GPAbook1}, DSs are excited in a dissipative system where a continuous supply of energy  is essential to keep the DSs alive. More specifically, in an active medium, the DSs are formed as a result of double balance between medium's gain and loss and between the GVD and nonlinearity. Theoretical investigation in this field has been growing over the years and the DSs are observed experimentally in amplifying optical medium such as, gain medium embedded waveguides and active fibers \cite{Akhmediev,Boardman,Xu,D-Li,Renninger}. In all the previous studies, it is considered that DSs evolve under positive nonlinearity, which is however not an essential criteria to excite DSs. In this work we try to explore the possibilities of forming DSs in different operational domain in terms of the numeric sign of dispersion and nonlinearity coefficient. Recent study has shown that ultrashort optical pulses experience frequency dependent nonlinearity in photonic crystal fibers (PCFs) doped with silver nanoparticles \cite{c18,bose,FRA-GPA} or waveguides employing quadratic nonlinear media \cite{JMoses}. The nonlinearity $\gamma$ of undoped PCFs is generally considered as constant and its first-order variation is taken care of by self-steepening term. In case of silver nanoparticle doped fibers the nonlinearity of composite system $\gamma_{eff}$ becomes a strong function of frequency (or wavelength) and even changes its numeric sign across zero-nonlinearity frequency (or wavelength), we generally call it zero-nonlinearity point (ZNP). Various interesting phenomena are observed as a consequence of ZNP, such as suppression and enhancement of Raman red-shifting of solitons and change of position and spectral bandwidth of dispersive waves (DWs) \cite{bose,FRA-GPA}. All the above analysis has been executed in conservative systems where external energy flow is prohibited. However, the idea of frequency dependent nonlinearity can be extended to the waveguides embedded in active gain medium or active fibers. The concept of negative nonlinearity in nonconservative system opens up new domain of DS excitation. Note, in conservative system, the standard Schr\"{o}dinger soliton comprises one or few parameter families due to its single balance between dispersion and nonlinearity. However, for dissipative system the soliton solutions are the result of the double balance and in general are isolated fixed points, where the pulse parameters are related to each other with fixed expressions. Such single soliton solution is named as Pereira-Stenflo (PS) soliton \cite{P-S}. The dynamics of DS has been investigated in both normal dispersion (ND) and anomalous dispersion (AD) domains where nonlinearity is positive (i.e., self-focussing) \cite{Xu,D-Li,Renninger}. However, the possibility of exciting DSs in self-defocusing or negative nonlinear media have never been explored. 

In this report, we try to investigate the formation and evolution of DSs inside an active medium whose nonlinearity can be self-focusing (positive) or self-defocusing (negative) type. Depending on the nature of dispersion we have four possible optical domains, 1 (ND, self-focusing nonlinearity); 2 (ND, self-defocusing nonlinearity); 3 (AD, self-defocusing nonlinearity) and 4 (AD, self-focusing nonlinearity). Stable solutions of the nonlinear Schr\"{o}dinger equation in the form of bright and dark solitons are obtained for AD and ND respectively when the nonlinearity is strictly positive. Unlike bright and dark Kerr solitons, the DSs are found in all four domains (including negative nonlinearity) followed by the governing equation in the form of a complex Ginzburg-Landau equation (GLE). We observe that, in domain 1 and 3 (where multiplication of GVD coefficient $\beta_2$ and nonlinear parameter $\gamma$ gives positive value) the DSs exhibit flattop spectra and are robust in nature. On the other hand, in domain 2 and 4 (where multiplications of $\beta_2$ and $\gamma$ gives negative value) the shape of the DS spectra exhibits standard $sech$ form. We exploit the variational treatment \cite{AS-SR,Bondeson, Kaup, Anderson, Cerda} to theoretically investigate the effects of physical perturbations like intra-pulse Raman scattering (IRS), self-steepening, third-order dispersion (TOD) and ZNP on the propagation of stable DS. The variational technique is a standard method used extensively in solving soliton problem in dissipative systems where the perturbations are included suitably into the Lagrangian density \cite{AS-SR}. The problem is further reduced by integrating the Lagrangian density followed by the Ritz optimization that leads to the equations of motion of individual pulse parameters under the perturbation. In our variational treatment we include the effect of the ZNP as a perturbation and investigate its effect. We further notice, the ZNP not only restricts or enhances the frequency down-shifting due to IRS, it also modifies the frequency location of TOD-mediated radiations. We theoretically investigate this phenomenon and establish a phase matching (PM) equation that predicts the spectral location of radiation frequency under ZNP.

We organize the work as follows; in Sec.~\ref{theory}, we introduce a realistic active waveguide that exhibits strong frequency dependent nonlinearity and can excite DS. In such a waveguide in addition to zero GVD wavelength we have zero-nonlinearity wavelength. We show that for such waveguide, based on the numeric signs of $\beta_2$ and $\gamma$, one can construct four operational regimes and DS can be excited in all four domains. In Sec.~\ref{perturbative}, we show the analytical results obtained from the variation analysis. In Sec.~\ref{RRS}, we study the impact of ZNP on IRS and correlate the numerical data with the variational results. Finally in Sec.~\ref{radiation}, we study how ZNP modifies the TOD-mediated radiation excited from DS \cite{AS-SR1}. We establish a modified PM equation that predicts the location of radiation as a function of the ZNP.

\section{The Ginzburg-Landau equation with negative nonlinearity}
\label{theory}
\begin{figure}[!htbp]
\begin{center} 
\epsfig{file=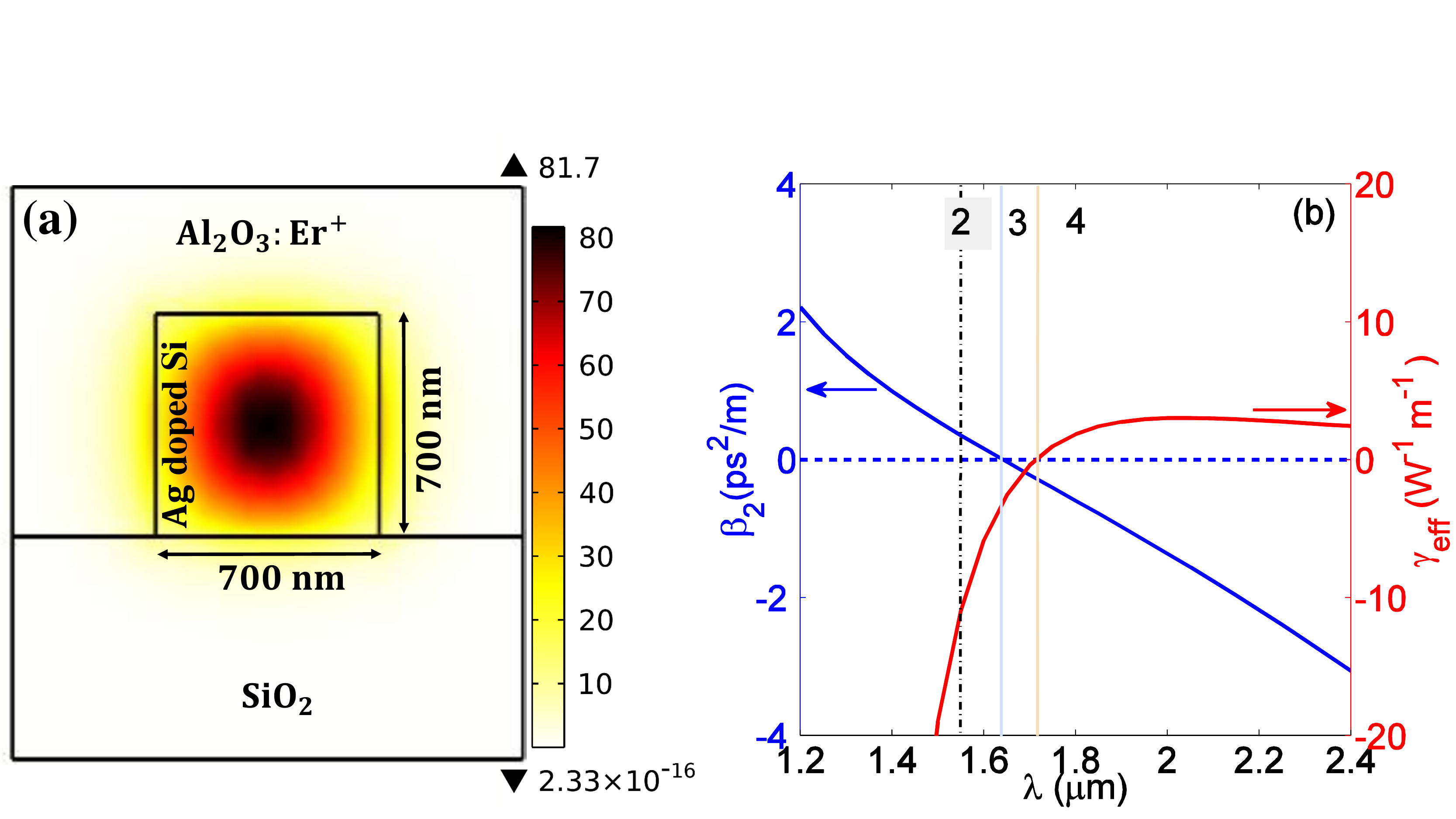,trim=0.0in 0.0in 0in 1.3in,clip=true, width=82mm} 
\caption{(a) Cross section of the proposed waveguide and transverse confinement of the quasi-TE mode is shown for $\lambda_0=1.55\, \mu m$ (domain 2) for a filling factor of $f=6\times 10^{-3}$. The width and height of the waveguide are $h=700\, nm$ and $w= 700 \,nm$, respectively. (b) The GVD and the nonlinear profiles are shown by the solid blue line and solid red line respectively. The location of the input wavelength ($\lambda_0=1.55\,\mu m$) is shown by the vertical dotted line. The domains 2, 3 and 4 are separated by the zero GVD and zero nonlinearity wavelengths.}  \label{fig:model}
\end{center}
\end{figure} 

\begin{figure*}[!htbp]
\begin{center} 
\epsfig{file=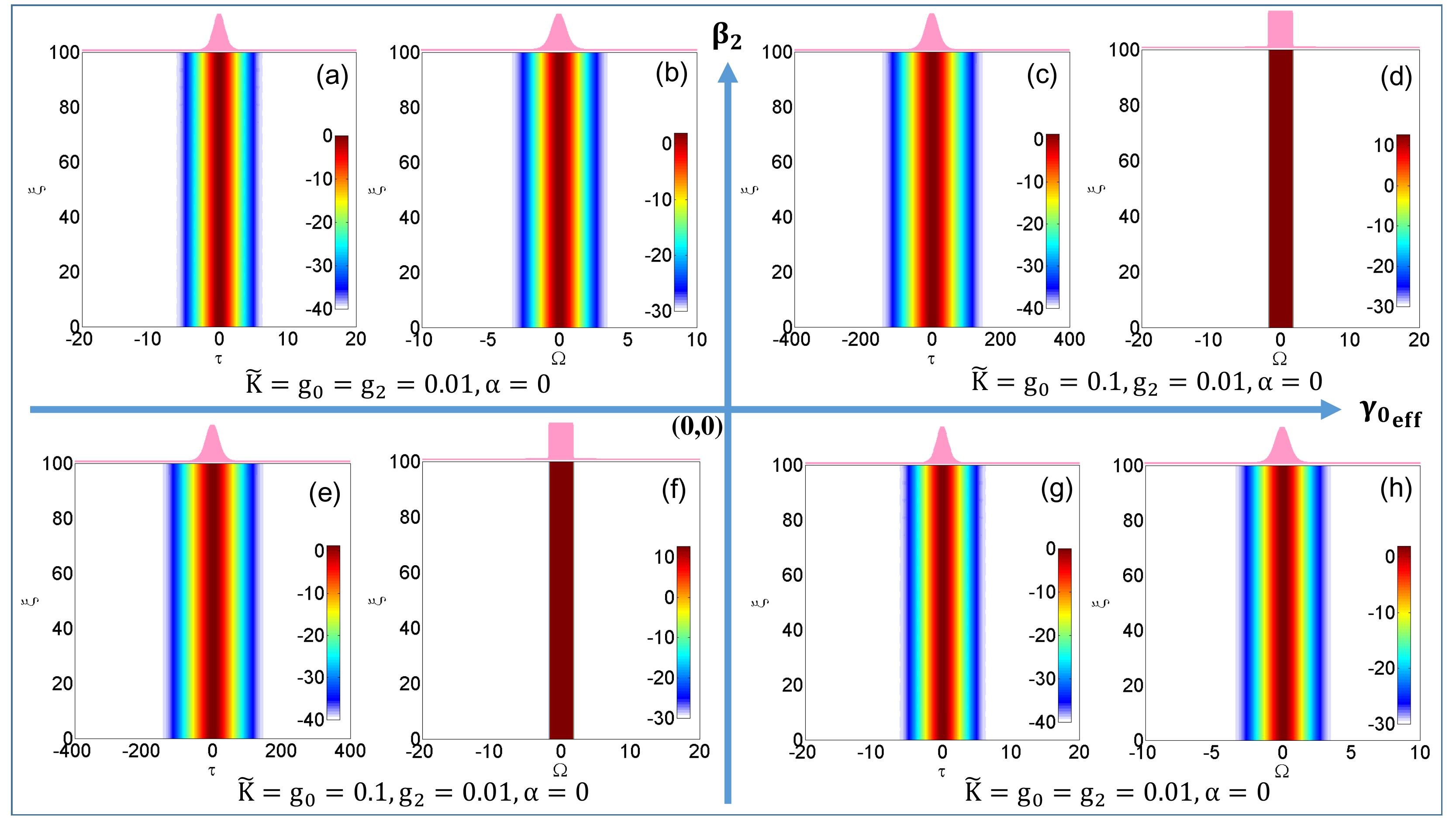,trim=0.0in 0.0in 0in 0in,clip=true, width=172mm}
\caption{Temporal and spectral dynamics of the exact solution Eq.\,\eqref{exact} of the unperturbed GLE in four different nonlinear dispersive domains. The evolution of DS in (a) Temporal and (b) spectral domain where the GVD is normal ($\beta_2>0$) and nonlinearity is self-defocussing type ($\gamma_{0eff}<0$) (domain 2). The same evolutions are described in (c)-(d); (e)-(f) and (g)-(h), where the GVD and nonlinearities are ($\beta_2>0$, $\gamma_{0eff}>0$); ($\beta_2<0$, $\gamma_{0eff}<0$) and ($\beta_2<0$, $\gamma_{0eff}>0$), and denoted by the domains 1; 3 and 4 respectively. Here $\Omega=2\pi(\nu-\nu_0)t_0$. The pulse shape on top of each panel gives the corresponding output intensity of stable DS.}  \label{fig:1}
\end{center}
\vspace{-0.6cm}
\end{figure*}

We propose an active Si-based waveguide shown in Fig.\,\ref{fig:model}(a) whose core is doped with silver nanoparticles. For this waveguide, the dispersion and nonlinear profiles are shown in Fig.\,\ref{fig:model}(b) for the wavelength range $1.2-2.4$ $\mu m$. By changing the geometry of the waveguide and density of doped nanoparticles we can tailor the positions of zero GVD wavelength and ZNP, which leads to four different operational domains. It is assumed that the doped nanoparticles solely influences the Kerr nonlinearity and the Raman contribution remains unaffected \cite{bose,FRA-GPA}. With this assumption, the GLE for optical field envelope $A(z,t)$ is written as,
\begin{align} \label{eq:19}
i\frac { \partial A }{ \partial z } +\sum _{ n=1 }^{ \infty  }{ \frac { { i }^{ n }{ \beta  }_{ n } }{ n! } \frac { { \partial  }^{ n }A }{ \partial { t }^{ n } }} 
-i\left(G +G_2 \frac{\partial^2 }{\partial t^2}  \right)A +i\alpha_l A \nonumber \\
+(1-f_R)\gamma_{eff}(\omega)|A(z,t)|^{ 2 } A(z,t)   
+ f_R \gamma(\omega)\nonumber \\ \times\left( A(z,t)\int _{ 0 }^{ \infty  }{ h_R(t')|A(z,t-t^{ ' })|^{ 2 }dt' }  \right) =0,
\end{align}
where $\beta_n$ is the $n^{th}$-order dispersion parameter and the nonlinear parameter of the undoped medium is $\gamma(\omega)=\gamma_R(\omega)+ i\gamma_I(\omega)$ with $\gamma_R(\omega)$ and $\gamma_I(\omega)$ are the real and imaginary parts respectively. The $\gamma_R(\omega)$ term gives the self-phase modulation effect and the $\gamma_I(\omega)$ gives the nonlinear absorption loss which, in our case, is the two-photon absorption (TPA) loss and assumed to be independent of $\omega$. $G$ and $G_2$ are the gain and gain dispersion coefficient, $\alpha_l$ is the linear loss coefficient and $f_R$ is the fractional Raman contribution. The term $\gamma_{eff}(\omega)$ is the nonlinear parameter of the doped medium. We can expand $\gamma_R(\omega)$ and $\gamma_{eff}(\omega)$ in the Taylor's series as: $\gamma_R(\omega)\approx \gamma_0 +\gamma_1 (\omega-\omega_0)$ and $\gamma_{eff}(\omega)\approx \gamma_0{_{eff}} +\gamma_1{_{eff}} (\omega-\omega_0)$. The parameters are defined as $\gamma_R (\omega )=\frac { 2\pi { n }_{ 2 }(\omega ) }{ { \lambda  }_{ 0 }{ A }_{ eff } }$, $\gamma_I=\frac{\beta_{TPA}}{2A_{eff}}$, $\gamma_{eff} (\omega )=\frac { 2\pi { n }_{ 2{eff} }(\omega ) }{ { \lambda  }_{ 0 }{ A }_{ eff } }$, ${{\gamma }_{1}}\approx \frac{{{\gamma }_{0}}}{{{\omega }_{0}}}$, ${{n}_{2}}=\frac{3\chi _{h}^{(3)}}{4{{\varepsilon }_{0}}n^2 c}$ and ${{n}_{2eff}}=\frac{3\chi _{eff}^{(3)}}{4{{\varepsilon}_{0}}c\,{{\varepsilon }_{eff}}}$, where $\lambda_0$ is the pump wavelength and $\beta_{TPA}$ is the TPA coefficient. Here ${{\varepsilon }_{eff}}=\frac{{{\varepsilon }_{h}}(1+2\sigma f)}{(1-\sigma f)}$ with $\sigma =\frac{{{\varepsilon }_{i}}-{{\varepsilon }_{h}}}{{{\varepsilon }_{i} +2{{\varepsilon }_{h}}}}$ is the effective permittivity of the silver-doped medium (in our case it is Si-based active waveguide) calculated from Maxwell-Garnett theory \cite{mgt}. The filling factor \textit{f} depends on the density of silver nanoparticles. Further, ${{\varepsilon }_{i}}$  and ${{\varepsilon }_{h}}$ are the dielectric functions of silver \cite{def} and Si, respectively. The effective third-order susceptibility of silver-doped Si waveguides has been calculated using the theory of composite nonlinear materials \cite{etna} as, 
\begin{equation} \label{eff}
\chi _{eff}^{(3)}=f\frac{{{\chi }^{(3)}_{i}}}{{{\left| B \right|}^{2}}{{B}^{2}}}+{{\chi }^{(3)}_{h}}\frac{D}{{{\left| 1-f\sigma  \right|}^{2}}{{(1-f\sigma )}^{2}}}
\end{equation}
with $B=\frac{(1-\sigma f)({{\varepsilon }_{i}}+2{{\varepsilon }_{h}})}{3{{\varepsilon }_{h}}}$ and $D=1-f\left\{ 1-0.4(4{{\sigma }^{2}}{{\left| \sigma  \right|}^{2}}+3\sigma {{\left| \sigma  \right|}^{2}}+{{\sigma }^{3}}+9{{\left| \sigma  \right|}^{2}}+9{{\sigma }^{2}}) \right\}.$ ${{\chi }^{(3)}_{h}}$ is the susceptibility of host Si. The susceptibility of silver is ${{\chi }^{(3)}_{i}}=(-6.3+i1.9)\times {{10}^{-16}}{{m}^{2}}/{{V}^{2}}$. The cross-section of the fundamental field distribution is shown in Fig.\,\ref{fig:model}(a). We use COMSOL software for a filling factor of $f=6\times10^{-3}$  to obtain the  GVD and nonlinear profiles. To study the impact of individual perturbations on a DS we reduce the integral form of GLE (Eq.\,\eqref{eq:19}) to a more convenient discrete form. The discrete form of the extended GLE (upto 1st order in $\gamma_{eff}$) is normalized as,
\begin{align}  \label{CGL}
i\frac{\partial u}{\partial \xi }&-\frac{1}{2}{\rm sgn}\left( {{\beta }_{2}} \right)\frac{{{\partial }^{2}}u}{\partial {{\tau }^{2}}}  -i\left( {{g}_{0}}+{{g}_{2}}\frac{{{\partial }^{2}}}{\partial {{\tau }^{2}}} \right)u+i\alpha u  \nonumber \\&\hspace{-0.5cm}
+ \left\{ {\rm sgn}(\gamma_{0eff})+i\,K f_R \right\}{{\left| u \right|}^{2}}u =i\delta_3 \frac{\partial ^3 u}{\partial \tau ^3} +\tau_R \Gamma_0  u \frac{\partial {\left| u \right|}^{2}}{\partial \tau } \nonumber \\ &\hspace{-0.5cm} +\left\{{\rm sgn}(\gamma_{0eff})- \Gamma_0\right\}f_R {\left| u \right|}^{2} u  
   -i\left\{\mu_{1eff}(1-f_R)\right. \nonumber \\ &\hspace{3.5cm}\left.+\Gamma_0 \,s f_R \right\} \frac{\partial{({\left| u \right|}^{2} u)}}{\partial \tau},
\end{align}
where $\Gamma_0=\gamma_0/\gamma_{0eff}$, $\mu_{1eff}=\gamma_{1eff}/\left(|\gamma_{0eff}|t_0\right)$. The time and space variables are normalized as $\tau=(t-zv_g^{-1})/t_0$ and $\xi=z/L_D$, where $v_g$ is the group velocity of the pulse, $t_0$ is the initial pulse width and $L_D=t_0^2/|\beta_2 (\omega_0)|$ is the dispersion length, $\beta_2(\omega_0)$ being the GVD coefficient at the carrier frequency  $\omega_0$. The TOD, IRS and self-stepping parameters are normalized as $\delta_3=\beta_3/(3!|\beta_2| t_0)$, $\tau_R=T_R/t_0$ and $s=1/(\omega_0 t_0)$, where $T_R$ is the first moment of the Raman response function \cite{GPAbook1}. The field amplitude $A$ is rescaled as, $A = u\sqrt{P_0}$, where peak power $P_0=|\beta_2 (\omega_0 )|/(t_0^2 \gamma_R)$. The dimensionless TPA coefficient is given as,  $K=\gamma_I/\gamma_R=\beta_{TPA}\lambda_0/(4\pi n_2)$.  The $\alpha_l$, $G$ and $G_2$ are normalized as $\alpha=\alpha_l L_D$, $g_0 = GL_D$ and $g_2= g_0(T_2/t_0)^2$ respectively, where dephasing time is $T_2$.  Note that, if we omit the right-side of Eq.\,\eqref{CGL} then it will readily reduce to the
well-known unperturbed GLE for which we have an exact solution of the form of PS soliton \cite{P-S}
\begin{equation} \label{exact}
u\left( \xi, \tau \right)={{u}_{0}}{{\left[ \text{sech}\left( \eta \tau  \right) \right]}^{\left( 1+ia \right)}}{{e}^{i\text{ }\!\!\Gamma\!\!\text{ }\xi }},
\end{equation}
where the four parameters $u_0,\ \eta,\ a$ and $\Gamma$ satisfy the following relations: 
\begin{align} \label{param}
|u_0|^{2}& = \frac{g_{0}-\alpha}{\widetilde{K} }\left[ 1-\frac{{\rm sgn}(\beta_{2})a/2 + g_{2}} {{g}_{2} ({{a}^{2}}-1 )- {\rm sgn}(\beta_2)a} \right], \nonumber \\ 
\eta^{2}& = \frac{{{g}_{0}}-\alpha}{ {{g}_{2}}\left( {{a}^{2}}-1 \right)-{\rm sgn}\left( {{\beta }_{2}} \right)a}, \nonumber \\
\Gamma &= \frac{{{\eta }^{2}}}{2}\left[ {\rm sgn}\left( {{\beta }_{2}} \right)\left( {{a}^{2}}-1 \right)+4a{{g}_{2}} \right], \nonumber \\
a &= \frac{H-{\rm sgn}(\gamma_{0eff})\sqrt{{{H}^{2}}+2{{\delta }^{2}}}}{\delta},
\end{align}
with $H=-[(3/2){\rm sgn}(\beta_2 )+3g_2 \widetilde{K}\,{\rm sgn}(\gamma_{0eff})]$,  $\delta=-[2g_2-{\rm sgn}(\beta_2) {\rm sgn}(\gamma_{0eff})\widetilde{K}]$ and $\widetilde{K}=K$, when the medium is undoped (nonlinearity is self-focusing type) and  $\widetilde{K}=K f_R$, when the medium is doped with silver nanoparticles (nonlinearity is self-defocusing or self-focusing type depending on the location of the ZNP). The numerical solution of unperturbed GLE (all the RHS terms\,$=0$ for Eq.\,\eqref{CGL}) is shown in Fig.\,\ref{fig:1}. Depending on the numeric signs of $\beta_2$ and $\gamma_{0eff}$ we can define four different regimes to excite DS. The numerical solution shows that the DSs are robust in all four domains and propagate without any distortion. In each case we launch pure PS soliton as input whose parameters obey the relationship shown in Eq.\,\eqref{param}. The full numerical solution reveals that, the same sign of the products of $\rm sgn(\beta_2)$ and $\rm sgn(\gamma_{0eff})$ give rise to identical structure of DS. In Fig.\,\ref{fig:1}, the set (a)-(b) has the identical structures as (g)-(h) (here, $\rm sgn{(\beta_2\times\gamma_{0eff})}=-1$ for both the cases). Similarly the set (c)-(d) and (e)-(f) are identical. For the second case, in Fig.\,\ref{fig:1} (d) and (f), a flat-top spectrum is observed in frequency domain of DS when both the GVD and nonlinearity have the same sign or their product is positive ($\rm sgn{(\beta_2\times \gamma_{0eff})}=+1$). Interestingly, the temporal width of such DS is found to be much wider compare to the previous case.

\section{VARIATIONAL ANALYSIS}
\label{perturbative}

The stable PS soliton evolves when the right-hand side of Eq.\,\eqref{CGL} vanishes. But it is interesting to study how the perturbations affect the stable PS soliton propagating under self-focusing and self-defocusing nonlinearity. To grasp the role of higher-order effects and negative nonlinearity, we study their impact through a variational analysis by treating them as small perturbations. The variation method is a standard technique which relies on the assumption that the overall shape of the input pulse remain unaffected during propagation. However, different pulse parameters like width, amplitude, phase, frequency, chirp etc. can vary. The choice of ansatz is an important step in such treatment \cite{GPAbook1}. In order to get the equation of motion of different pulse parameters, we first write Eq.~\eqref{CGL} in the form of a perturbed nonlinear Schr\"{o}dinger equation:\cite{Anderson, GPAbook1}
\begin{equation} \label{nls1}
    i\frac{\partial u}{\partial\xi}-\frac{1}{2}p\frac{\partial^{2}u}
    {\partial\tau^{2}}+ q|u|^{2}u = i\epsilon(u),
\end{equation}
where $p={\rm sgn}(\beta_2$), $q={\rm sgn}(\gamma_{0eff})$ and  the small perturbation $\epsilon(u)$ is defined as,
\begin{align} \label{nls2}
    \epsilon(u)& = \delta_3\frac{\partial^3u}{\partial{{\tau}^{3}}}-i\tau_R \Gamma_0  u \frac{\partial{|u|}^{2}}{\partial\tau} -\left\{\mu_{1eff}(1-f_R)\right. \nonumber \\ &\hspace{0.5cm}\left.+\Gamma_0 s f_R \right\}\frac{\partial ({|u |}^{2} u)}{\partial \tau}  -i(q-\Gamma_0)f_R |u|^{2}u \nonumber \\& \hspace{2cm}
 -Kf_R|u|^{2}u +{g}_{0}u +{{g}_{2}}\frac{{{\partial }^{2}u}}{\partial {{\tau }^{2}}} .
\end{align}
For this specific problem, we choose the PS soliton (see Eq.\,\eqref{exact}) as our ansatz since it is the exact solution of Eq.\,\eqref{CGL} in absence of all the perturbations. The complete mathematical form of the ansatz is as follows,
\begin{align}\label{nls0}
 u\left( \xi ,~\tau  \right)&={{u}_{0}}\left( \xi  \right){{\left( \text{sech}\left[ \eta \left( \xi  \right)\left\{ \tau -{{\tau }_{p}}\left( \xi  \right) \right\} \right] \right)}^{\left\{ 1+ia\left( \xi  \right) \right\}}} \nonumber\\&\hspace{0.5cm}
\exp\left( i\left[ \phi \left( \xi  \right)-\text{ }\Omega_p\left( \xi  \right)\left\{ \tau -{{\tau }_{p}}\left( \xi  \right) \right\} \right]  \right).
\end{align}
Note, the six parameters $u_0$, $\eta$, $\tau_p$, $\phi$, $a$ and $\Omega_p$ are now assumed to be a function of $\xi$. 
Next we introduce a Lagrangian density appropriate for Eq.\,\eqref{nls1} as, $\mathcal{L_D}=\frac{i}{2}(u u_{\xi}^* -u^* u_\xi )-\frac{1}{2}\left(q|u|^4+p|u_\tau|^2 \right)$. We integrate $\mathcal{L_D}$ over $\tau$ using the ansatz given in Eq.\,\eqref{nls0} and obtain the following reduced Lagrangian:
\begin{gather}
    L = \int_{-\infty }^{\infty}\mathcal{L_D} d\tau = \frac{2{{u}_{0}}^{2}}{\eta }\left( \frac{\partial \phi }{\partial \xi }+\Omega_p \frac{\partial {{\tau }_{p}}}{\partial \xi } \right)-\frac{a {{u}_{0}}^{2}}{{{\eta }^{2}}}\frac{\partial \eta }{\partial \xi }\nonumber\\+C\frac{{{u}_{0}}^{2}}{\eta }\frac{\partial a}{\partial \xi }  
    -p\frac{\eta {{u}_{0}}^{2}}{3}\left( 1+{{a}^{2}} \right) -\frac{{{u}_{0}}^{2}}{\eta }\left( p\,{{\Omega_p }^{2}}+q\frac{2}{3}{{u}_{0}}^{2} \right) \nonumber \\
    +i\int_{-\infty }^{\infty}(\epsilon{u}^{*}-\epsilon^* u)\,d\tau,
\end{gather}
where $C=[\ln(4)-2]$. Finally we use the Euler-Lagrange equation for each pulse parameter to obtain a set of coupled ODEs for the six parameters that describe the overall soliton dynamics \cite{GPAbook1,Hasegawa-K}. These equations describe the evolution of pulse energy $(E=\int_{-\infty}^{\infty} |u|^{2}\,d\tau)$, temporal position $\tau_p$, frequency shift $\Omega_p$, amplitude $\eta$, frequency chirp $a$, and phase $\phi$ with the following form,

\begin{align}
  \frac{dE}{d\xi }&=\frac{d}{d\xi }\left( \frac{2{{u}_{0}}^{2}}{\eta} \right) =2{\rm Re}\int\limits_{-\infty }^{\infty }{\epsilon {{u}^{*}}}\,d\tau, \label{var1}  \\
    \frac{d{{\tau}_{p}}}{d\xi }& = p\,\Omega_p +\frac{\eta }{{{u}_{0}}^{2}} \int\limits_{-\infty }^{\infty }{\left( \tau -{{\tau }_{p}} \right){\rm Re}\left( \epsilon {{u}^{*}} \right)\,d\tau }, \label{var2}\\
    \frac{d\Omega_p }{d\xi}&= \frac{{{\eta }^{2}}}{{{u}_{0}}^{2}}\int\limits_{-\infty }^{\infty }{\tanh\left[ \eta \left( \tau -{{\tau }_{p}} \right) \right]{\rm Re}\left[ \left( a+i \right)\epsilon {{u}^{*}} \right]\,d\tau }, \label{var3}
\end{align}
\begin{align}
    \frac{d\eta }{d\xi}&=  -C\frac{\eta^{2}}{2u_{0}^{2}}E_\xi - p\frac{2{{\eta}^{3}}a}{3} +\frac{2{{\eta }^{2}}}{{{u}_{0}}^{2}}\nonumber \\ &\times \int\limits_{-\infty }^{\infty }{\ln\left[ {\rm sech}\left\{ \eta \left( \tau -{{\tau }_{p}} \right) \right\} \right]{\rm Re}\left( \epsilon {{u}^{*}} \right)\,d\tau }, \label{var4}\\
\frac{da}{d\xi} &=-\frac{2{{\eta }^{2}}}{{{u}_{0}}^{2}}{\rm Im}\int\limits_{-\infty }^{\infty }{\left( \tau -{{\tau }_{p}} \right)\tanh\left[ \eta \left( \tau -{{\tau }_{p}} \right) \right]\left( 1-ia \right)\epsilon {{u}^{*}}\,d\tau } \nonumber \\
&\hspace{-0.5cm}-\frac{a\eta }{2{{u}_{0}}^{2}}E_\xi +q\frac{2}{3}{{u}_{0}}^{2} +p\frac{2}{3}{{\eta }^{2}}\left( 1+{{a}^{2}} \right) +\frac{\eta }{{{u}_{0}}^{2}}{\rm Im}\int\limits_{-\infty }^{\infty }{\epsilon {{u}^{*}}}\,d\tau , \label{var5}\\
\frac{d\phi}{d\xi} &= \frac{a}{2\eta}\eta_\xi -\frac{C}{2}a_\xi  -\Omega_p \,{\tau_p}_\xi + \frac{2}{3}q\,u_0^2 +\frac{1}{6}p\,\eta^2(1+a^2) \nonumber \\&\hspace{2cm}+\frac{1}{2}p\,\Omega_p^2    +\frac{\eta}{2u_0^2}{\rm Im}\int\limits_{-\infty }^{\infty}{\epsilon {{u}^{*}}}\,d\tau, \label{var6}
\end{align}
where Re and Im stand for real and imaginary parts. The final step is to evaluate all the integrals using $\epsilon(u)$ given in Eq.\ \eqref{nls2}. It results in the following set of six coupled differential equations:
\begin{align}
\frac{dE}{d\xi}&=\frac{2}{3}(3g_0-Kf_R\eta E)E- \frac{2}{3}g_2[(1+a^2)\eta^2+3\Omega_p^2]E, \label{var7} \\
\frac{d{{\tau }_{p}}}{d\xi}&=\left(p -2{{g}_{2}}a \right)\Omega_p +\frac{1}{2}\left\{\mu_{1eff}(1-f_R) +\Gamma_0 s f_R \right\} \eta E \nonumber \\ &\hspace{3cm}+\delta_3\left[(1+a^2)\eta^2 + 3\Omega_p^2 \right], \label{var8}\\
\frac{d\Omega_p }{d\xi}&=-\frac{4}{3}{{g}_{2}}\left( 1+{{a}^{2}} \right)\Omega_p {{\eta }^{2}}   -\frac{4}{15}\tau_R \,\Gamma_0 E{{\eta }^{3}} \nonumber \\ &\hspace{1cm}+ \frac{4}{15}\left\{\mu_{1eff}(1-f_R) +\Gamma_0 s f_R \right\} a E \eta^3, \label{var9}\\
\frac{d\eta }{d\xi }&=-\frac{2}{9} (3p\,a\eta + EKf_R)\eta^2-\frac{4}{9}(2-a^2)g_2\eta^3 \nonumber \\&\hspace{4.5cm}-4\delta_3 a \Omega_p \eta^3,\label{var10}\\
\frac{da}{d\xi }&=\frac{1}{3}\{q(1-f_R) +(Ka+\Gamma_0)f_R \}E\eta \nonumber \\ &\hspace{1cm}+\frac{2}{3} (p-g_2a)(1+a^2)\eta^2 +4\delta_3 \Omega_p \eta^2(1+ a^2) \nonumber \\& \hspace{1.2cm}+ \frac{1}{3}\left\{\mu_{1eff}(1-f_R) +\Gamma_0 s f_R \right\} \Omega_p \eta E , \label{var11}\\
\frac{d\phi}{d\xi}&=\frac{a}{2\eta}\eta_\xi -\frac{C}{2}a_\xi  -\Omega_p{\tau_p}_\xi + \frac{1}{3}q\,\eta E +\frac{1}{6}p\,\eta^2(1+a^2)\nonumber \\&\hspace{0.5cm}+\frac{1}{2}p\,\Omega_p^2 +\frac{1}{3}\left\{\mu_{1eff}(1-f_R) +\Gamma_0 s f_R \right\}\Omega_p E \eta \nonumber \\&\hspace{0.0cm}-\frac{1}{3}(q-\Gamma_0)f_RE\eta+ \delta_3 \left[(1+a^2)\Omega_p \eta^2 + \Omega_p^3 \right].
\label{var12}
\end{align}
These equations are quite significant to understand the pulse dynamics. Each equation provides considerable physical insight as they show how a perturbation affects a specific pulse parameter. For example, the Raman parameter $\tau_R$ appears only in the equation for the frequency shift $\Omega_p$ and the term containing it has a negative sign. This immediately shows that the IRS leads to a spectral redshift of the DS. The term $\mu_{1eff}$ appears in the frequency equation which modifies the self-steepening parameter $s$ and depending on the signs of $\mu_{1eff}$ the Raman redshift will be restricted (for $\mu_{1eff}<0$) or relaxed more (for $\mu_{1eff}>0$). The presence of the term $\mu_{1eff}$ in the equations of temporal position $\tau_p$ and frequency chirp $a$  will also modify the dynamics of DS. 

\section{Influence of the ZNP on Raman redshift}
\label{RRS}
\begin{figure}[tb!]
\begin{center}
\epsfig{file=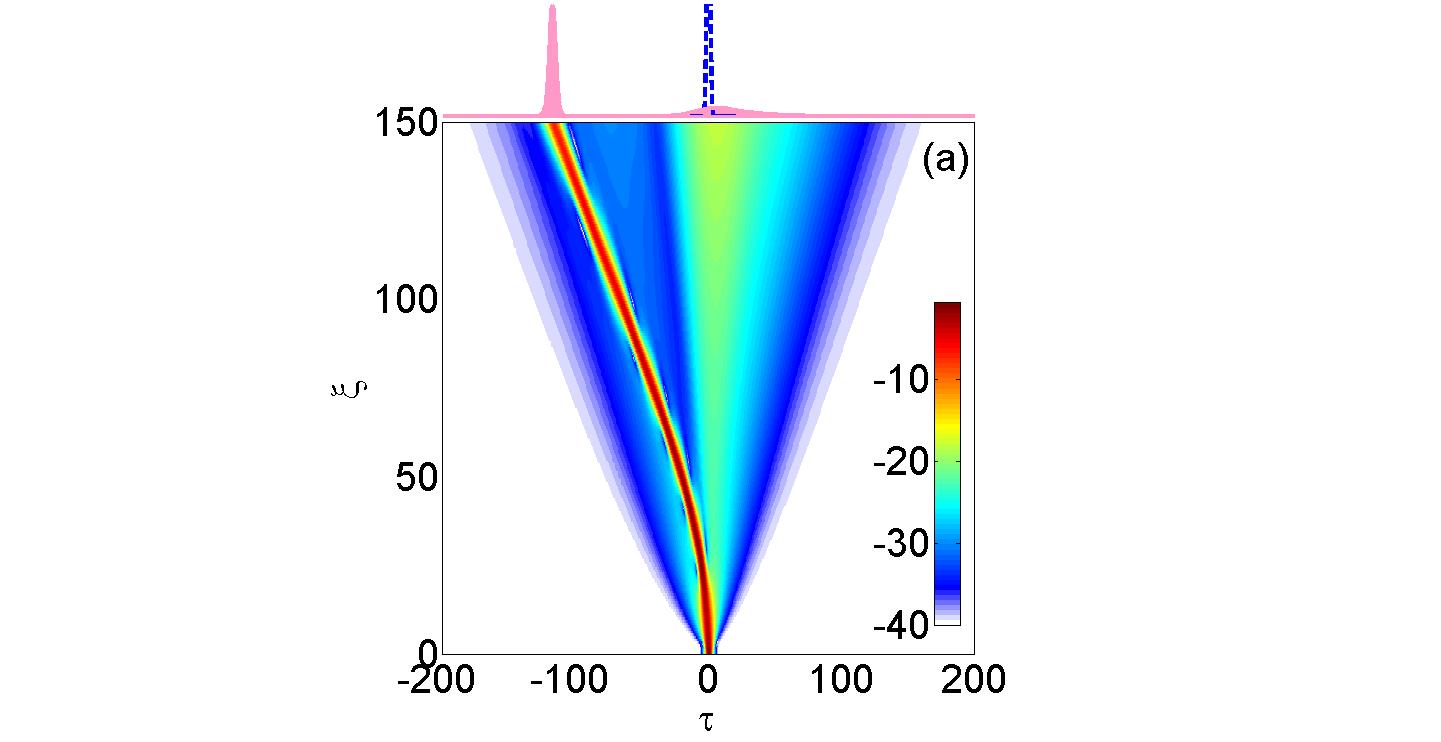,trim=4.5in 0.05in 6.05in 0.0in,clip=true, width=42mm}
\epsfig{file=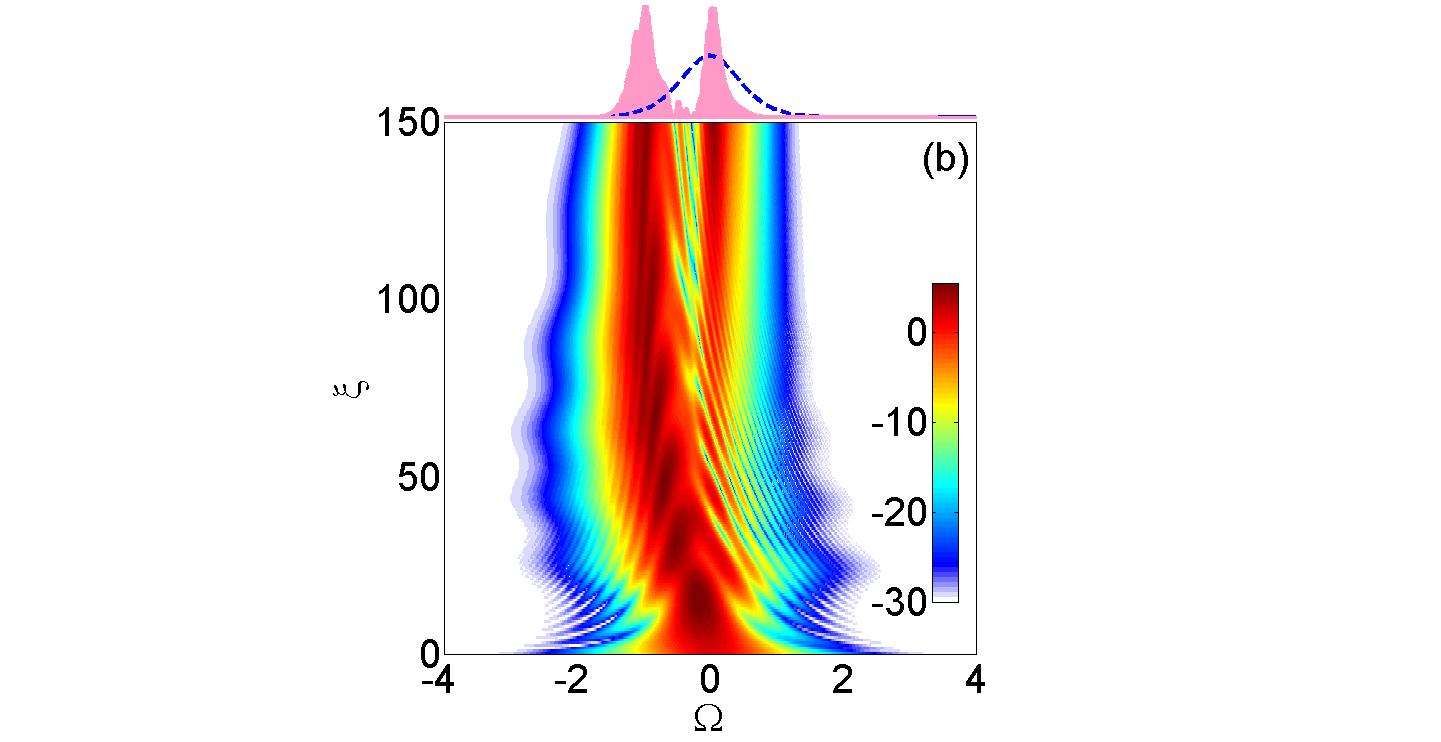,trim=4.5in 0.05in 6.05in 0.0in,clip=true, width=42mm}
\vspace{0.0em}
\epsfig{file=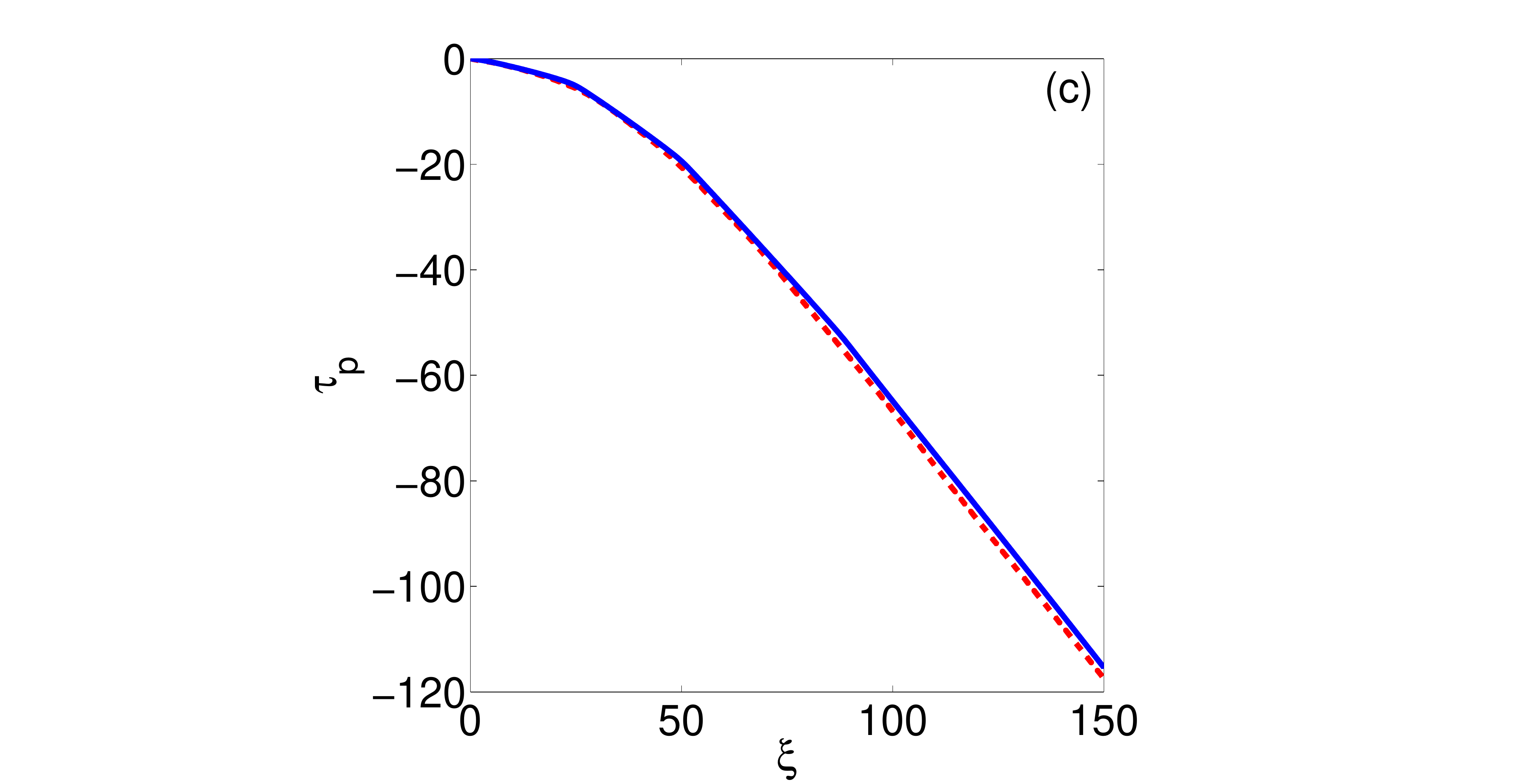,trim=3in -0.3in 3.75in 0.4in,clip=true, width=42mm}
\epsfig{file=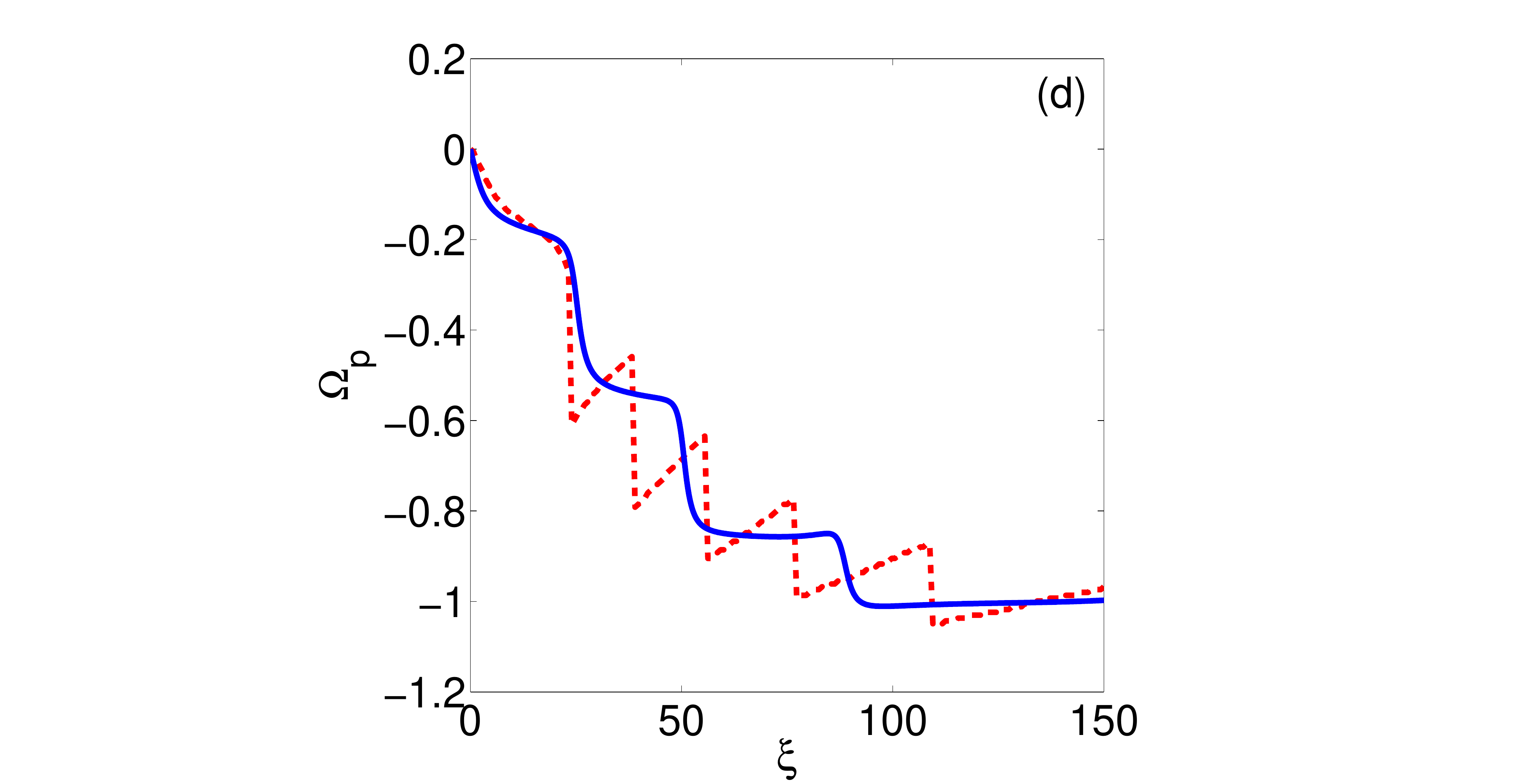,trim=3.0in -0.3in 3.75in 0.4in,clip=true, width=42mm}
\vspace{-1.0em}
\caption{(Color online) (a) Temporal and (b) spectral evolution of the perturbed DS in domain 2 ($\beta_2>0$ and $\gamma_{0eff}<0$) for the parameters  $K=0.05,~g_0=g_2=0.01,~\Gamma_0=0.2,f_R=0.245,~\tau_R=0.5,~s=0.1$ and $\mu_{1eff}=-0.1$. The input (dotted trace) and output pulse shapes are also shown in the top panel. Variation of (c) temporal peak position shift ($\tau_p$) and (d) frequency position shift ($\Omega_p$) over the propagation distance. The solid lines give the variational results, whereas the dotted lines give the numerically simulated results.}\label{variational_all}
\end{center}
\vspace{-0.7cm}
\end{figure}

\begin{figure*}[tb!]
\begin{center}
\epsfig{file=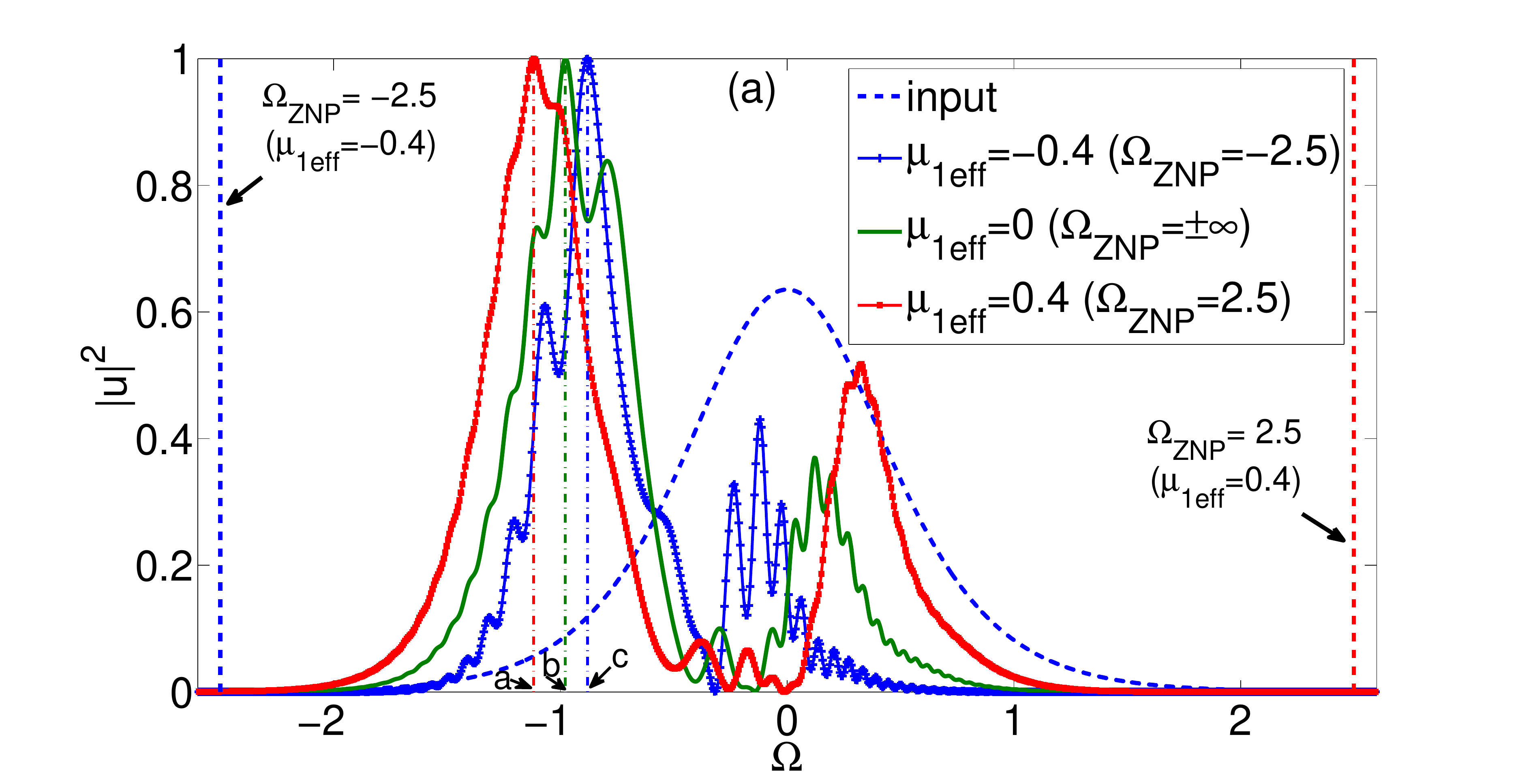,trim=0.7in 0.1in 1.3in 0.4in,clip=true, width=88mm}
\epsfig{file=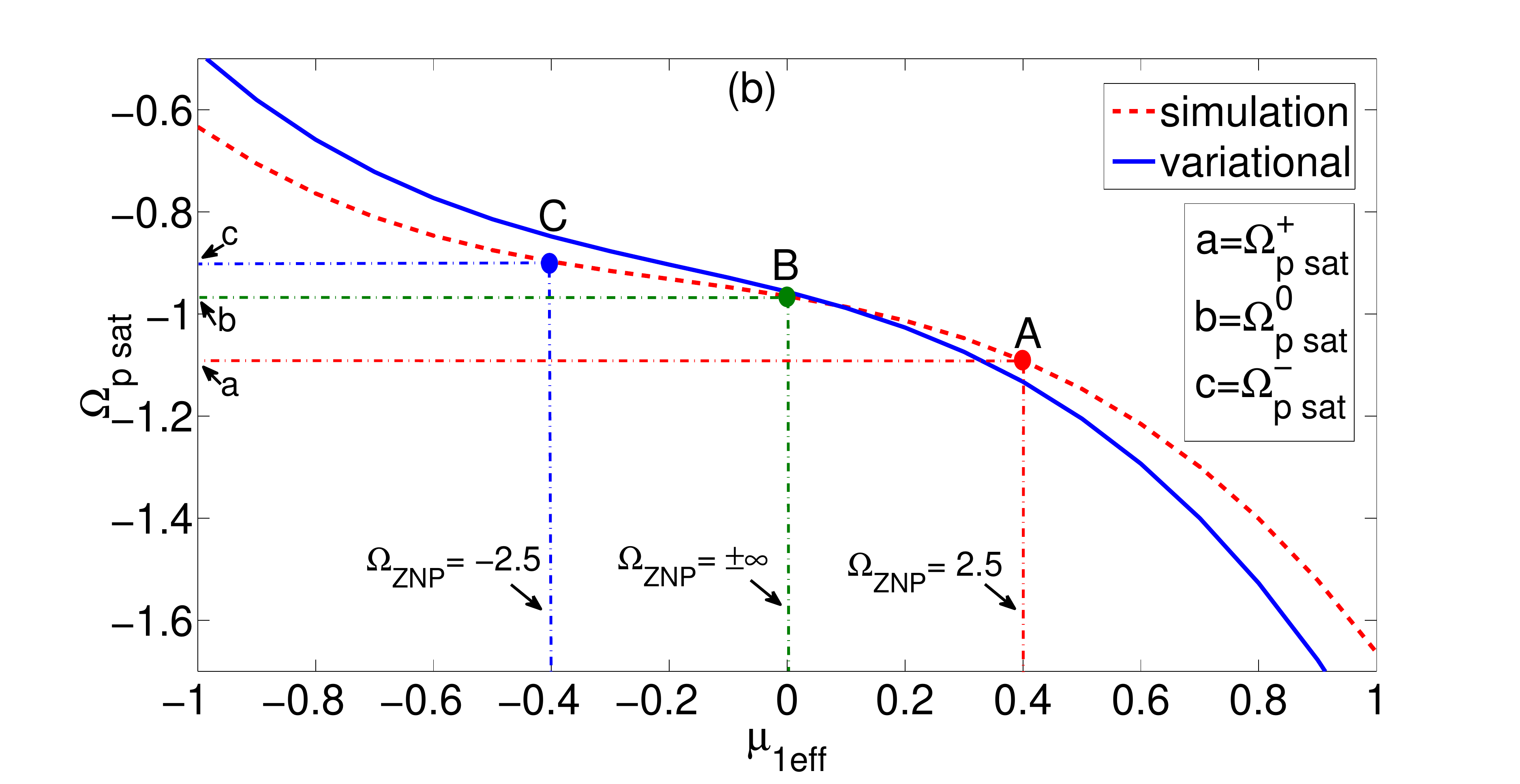,trim=0.4in 0.1in 1.3in 0.4in,clip=true, width=90mm}
\vspace{-2em}
\caption{(Color online) (a) Raman induced spectral redshift for $\mu_{1eff}=0,\pm0.4$ at $\xi=100$. (b) $\Omega_{p~sat}$ as a function of $\mu_{1eff}$ plot. 
The parameters used in the simulations are the same as Fig.\,\ref{variational_all}. }\label{fig4all}
\end{center}
\vspace{-0.6cm}
\end{figure*}

The impact of ZNP is significant on Raman redshift. Suppression of Raman red-shifting due to two-zero dispersion is a well-known phenomenon in Kerr solitons where strong dispersive waves are generated across the second zero GVD point \cite{DVS-PSJR}. Recently, it is observed that the Raman red-shifting of Schr\"{o}dinger Kerr soliton is also restricted in the presence of ZNP \cite{bose,FRA-GPA}. However, this suppressed Raman frequency shift does not lead to any other radiation across ZNP. In other words, in this case the ZNP acts as a perfect barrier that not
only restricts the Raman redshift but also does not allow any radiation beyond this point. The earlier reported works \cite{bose,FRA-GPA} on ZNP are mostly based on numerical observations with less theoretical explanation. Here we try to address this specific problem theoretically for DSs which experiences the ZNP. Like Kerr solitons, the impact of IRS term is characteristically same for DSs where we get frequency red-shifting. To study the impact of IRS, we exclude the TOD term (i.e., $\delta_3=0$) and numerically solve the extended GLE Eq.\,\eqref{CGL} using the standard split-step Fourier method \cite{GPAbook1}. The input pulse is taken in the form of a PS soliton whose parameters obey the relations shown in Eq.\,\eqref{param}. Finally the full numerical results are compared with the results obtained from the variational method by solving the set of coupled ordinary differential equations Eqs.\,\eqref{var7}-\eqref{var11}. Figures.\,\ref{variational_all}(a) and (b) show the density plots of temporal and spectral evolution of an optical DS under IRS. The temporal position ($\tau_p$) and frequency shift ($\Omega_p$) with respect to the propagation distance are also plotted in Figs.\,\ref{variational_all}(c) and (d) respectively, where results from variational analysis and full numerical simulation are superimposed. The variational results (solid lines) corroborate well with the full numerical results (dotted lines). Next, we try to understand the influence of ZNP on Raman redshift. The slope of the nonlinear dispersion curve is mathematically defined by the parameter $\mu_{1eff}$. Note that, the value of the slope ($\mu_{1eff}$) determines the location of the ZNP. In Fig.\,\ref{fig4all}, we plot the output spectrum for three different cases where the values of $\mu_{1eff}=0,\pm0.4$. When $\mu_{1eff}=0$, the nonlinear dispersion curve is independent of frequency and does not contain any ZNP. In such case, the DS accelerates freely and its spectrum is shown by the solid green line in Fig.\,\ref{fig4all}(a). Now, the Raman redshift is reduced significantly when a ZNP is introduced with a negative slope ($\mu_{1eff}=-0.4$). The location of the ZNP is shown by a vertical blue dotted line and it behaves like a barrier that restricts the frequency down-shifting. Interestingly, the redshift is enhanced (red line) when a ZNP is located beyond the launching point in positive frequency side (denoted by a vertical red dotted line). In order to understand this phenomena quantitatively, we exploit the variational results. It is observed that, the frequency red-shifting due to IRS is saturated over the distance. This saturated frequency is denoted by $\Omega_{p\,sat}$. The full variational treatment can predict $\Omega_{p\,sat}$, which is a function of $\mu_{1eff}$. In Fig.\,\ref{fig4all}(a), the peak frequency positions are denoted by $a=\Omega_{p\,sat}^+$ (for $\mu_{1eff}=+0.4$), $b=\Omega_{p\,sat}^0$ (for $\mu_{1eff}=0$) and $c=\Omega_{p\,sat}^-$ (for $\mu_{1eff}=-0.4$). In Fig.\,\ref{fig4all}(b) we try to map the results obtained in Fig. 4(a), where we use the simulated values to plot $\Omega_{p\,sat}$ as a function of $\mu_{1eff}$. In Fig.\,\ref{fig4all}(b), we also locate the three points $A,~B$ and $C$ for $\mu_{1eff}=+0.4,~0$ and $-0.4$, respectively. We superimpose the variational result (solid blue line) in the same plot, and find a close agreement between them. The physical origin of suppressed and enhanced Raman redshift can be explained from asymmetric pulse broadening due to frequency dependent nonlinearity \cite{GPAbook1}. The term $\mu_{1eff}$ acts as a modulation to the self-steepening strength and even changes its sign. When $\mu_{1eff}>0\,(<0)$, it increases (decreases) the total self-steepening strength and under self-defocussing nonlinearity, it gives a small frequency shift towards lower (higher) frequency side, and this effect results in enhanced (suppressed) Raman redshift combining with IRS perturbation. This is also validated from variational method by the Eq.\,\eqref{var9}.

\section{Dissipative soliton mediated radiation under the ZNP}
\label{radiation}
\begin{figure}[!htbp]
\begin{center}
\epsfig{file=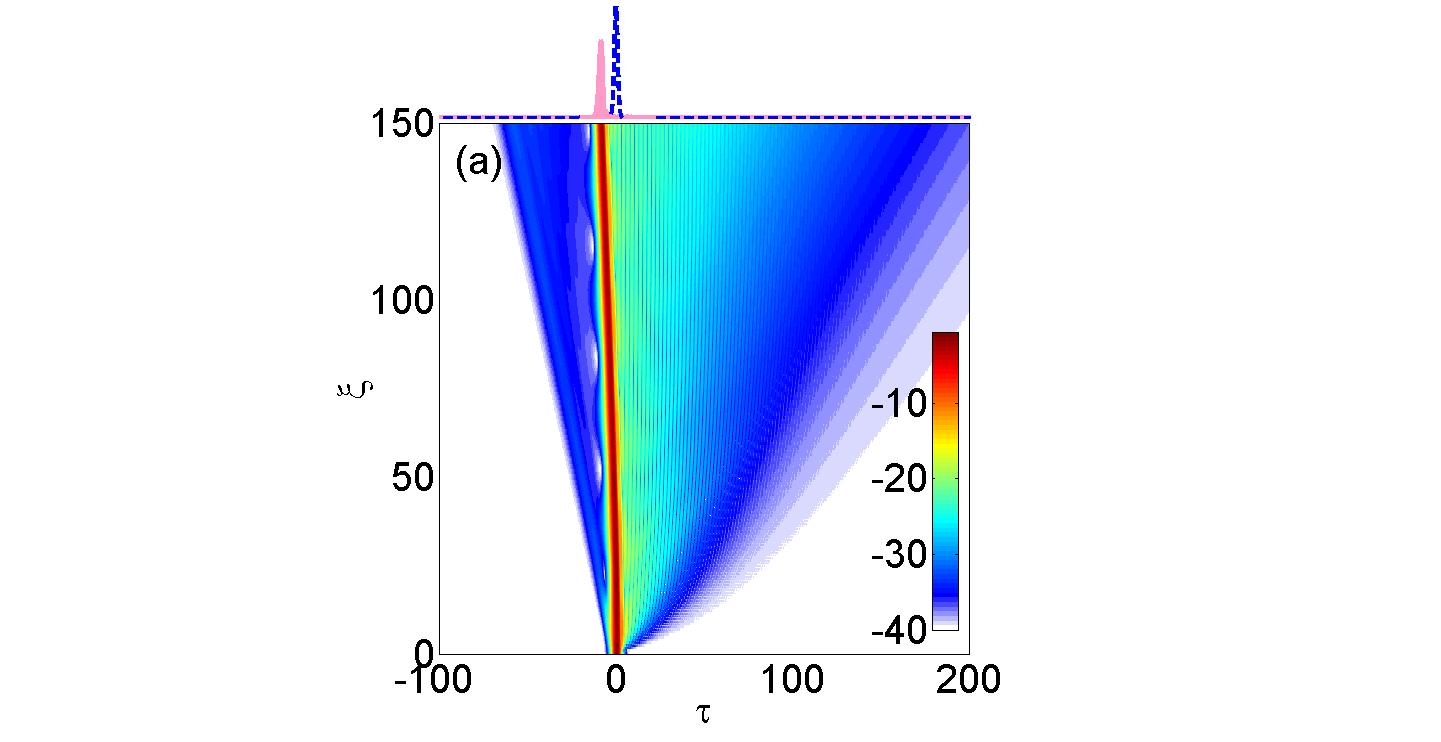,trim=4.5in 0.05in 6.05in 0.0in,clip=true, width=41mm}
\hspace{0.3em}
\epsfig{file=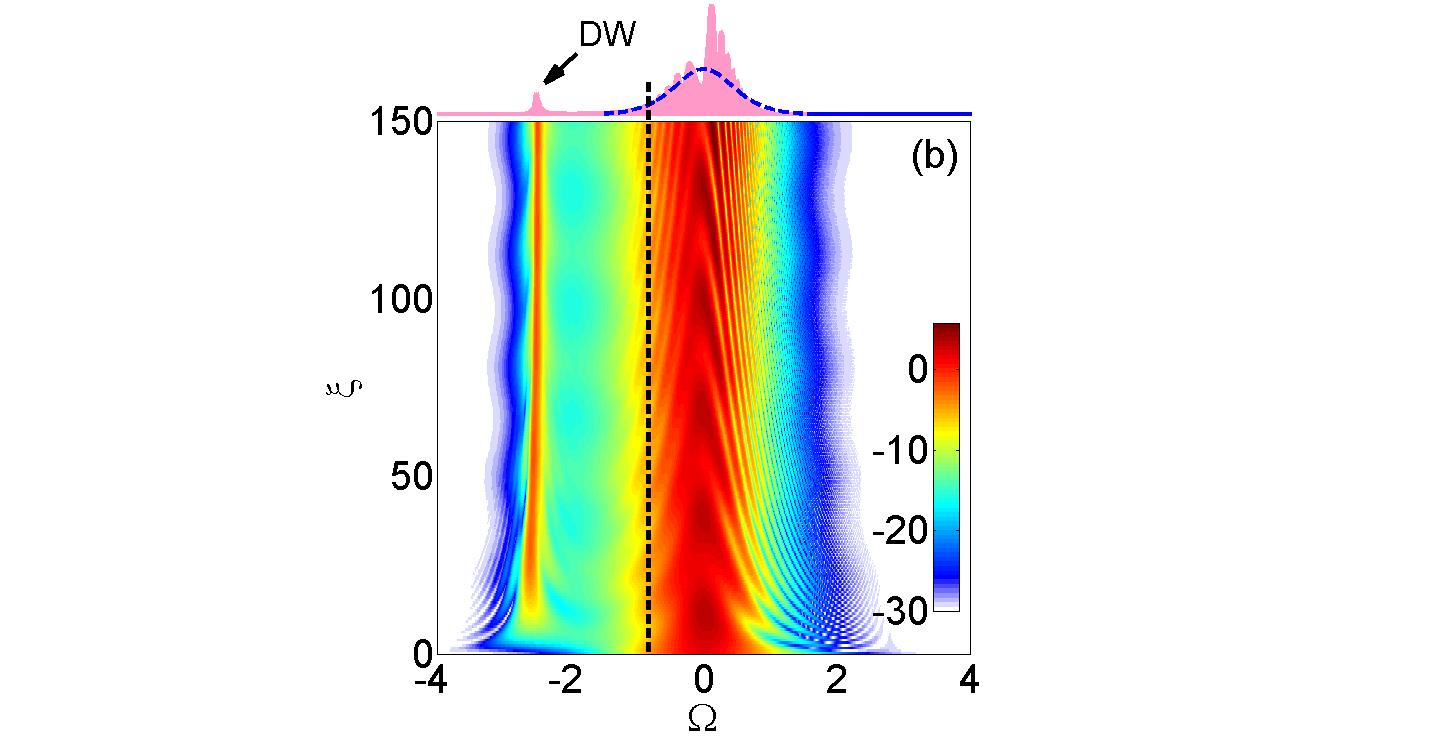,trim=4.5in 0.05in 6.05in 0.0in,clip=true, width=41mm}
\vspace{-0.0em}
\epsfig{file=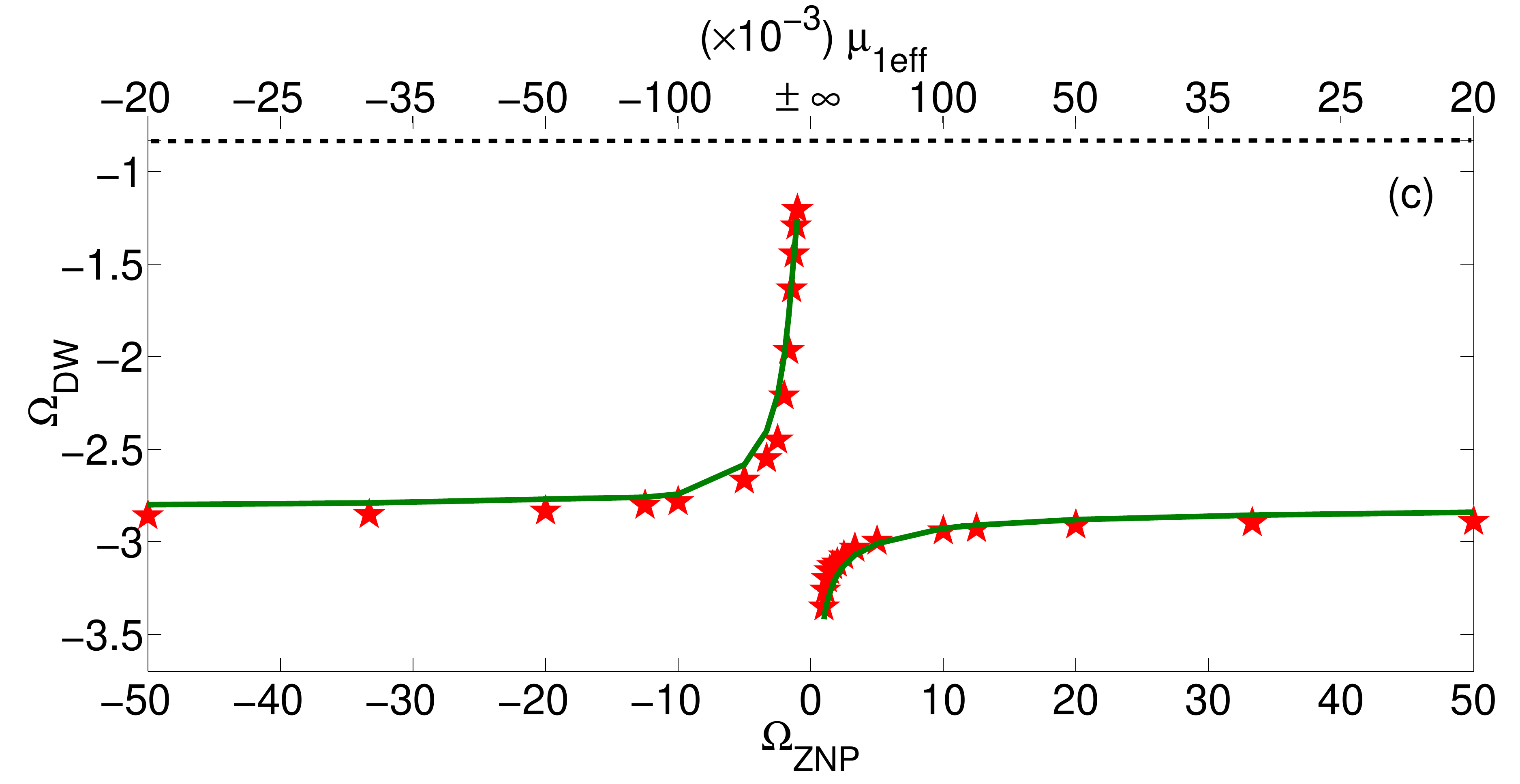,trim=0.25in 0.00in 0.2in 0.0in,clip=true, width=84mm}
\vspace{0.0em}
\epsfig{file=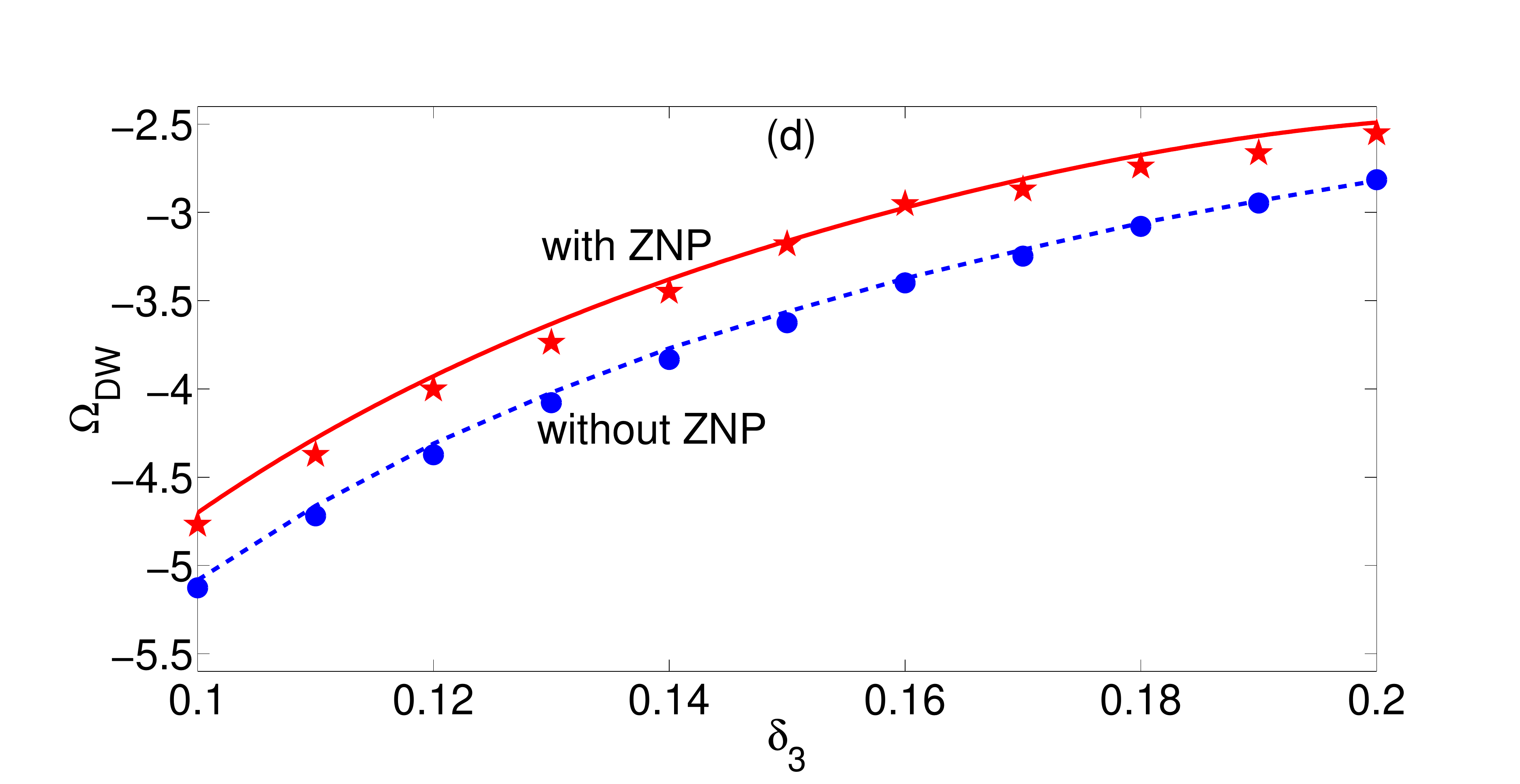,trim=0.7in 0.05in 1.1in 1.0in,clip=true, width=84mm}
\vspace{-1.3em}
\caption{(Color online)(a) Temporal and (b) spectral evolution of DS in presence of TOD ($\delta_3=0.2$) and ZNP ($\mu_{1eff}=-0.1$) along with all other perturbation values same as Fig.\,\ref{variational_all}. The input (dotted trace) and output pulse shapes are also shown in the top panel.
(c) The variation of the frequency location of DW ($\Omega_{DW}$) with the variation of ZNP ($\mu_{1eff}$). The solid lines represent the results obtained from the PM equation Eq.\,\eqref{PM}, whereas the stars give the numerically simulated results. The vertical and horizontal dotted lines in (b) and (c) represent the locations of zero GVD position respectively. (d) The variation of $\Omega_{DW}$ with TOD parameter in the presence of ZNP ($\mu_{1eff}=-0.3$) and in the absence ZNP ($\mu_{1eff}=0$). The stars (circles) give the simulated results, whereas the solid red line (dashed blue line) gives the prediction from the PM equation, when the ZNP is present (absent).
}\label{variational_all_TOD}
\end{center}
\vspace{-0.6cm}
\end{figure}

The TOD is the primary perturbation that a soliton experiences when it propagates through a dispersive medium. Being an asymmetric term, TOD leads to some temporal side-lobes and is responsible for generating a phase-matched radiation in the form of a DW \cite{NA-MK}. In one of our previous works, it has already been shown that like Kerr soliton, the DS can also radiate DW in the vicinity of zero GVD wavelength \cite{AS-SR1}. However, the situation is rather interesting when DS experiences a ZNP in addition to the zero GVD wavelength. The relative location of the zero GVD wavelength and ZNP compared to the input wavelength significantly influences the dynamics of the DS. We solve the GLE numerically considering TOD as perturbation, and the simulated results are shown in Fig.\,\ref{variational_all_TOD}. The formation of DW across the zero GVD point (shown by a vertical dotted line) is evident. We run our simulation in presence and in absence of ZNP where we find the frequency location of DW is modified by the ZNP. We introduce the effect of ZNP in the PM equation and derive the following equation in normalised unit \cite{AS-SR1},
\begin{align} \label{PM}
\frac{sgn(\beta_2)}{2}\Omega^2 + \delta_3 \Omega^3 -\tau_g \Omega= 2\Gamma + sgn(\gamma_{0eff})\Gamma \, \mu_{1eff}\Omega. 
\end{align}
Here, $\tau_g=(v^{-1}_{DW}-v^{-1}_{DS}) t_0 |\beta_2|^{-1}$, is the normalized form of the inverse-velocity mismatch between the DS and DW. The DS ansatz parameter $\Gamma$ takes the form:  $\Gamma\approx \frac{1}{2}sgn(\gamma_{0eff})u_0^2$. The solution ($\Omega_{DW}$) of Eq.\,\eqref{PM} provides the radiation frequency of the DW. It is interesting to note that, unlike zero GVD point, ZNP does not lead to any phase-matched radiation. It can be verified from Eq.\,\eqref{PM} that, it does not produce any real solution for the given range of $\mu_{1eff}$ when $\delta_3=0$. In Figs.\,\ref{variational_all_TOD} (a) and (b), we demonstrate the temporal and spectral evolution of DS in presence of the TOD ($\delta_3=0.2$) for $\gamma_{0eff}<0$. To get the complete picture, we include all the perturbations (Raman, self-steeping, etc.) in the simulation. It is observed that the spectral position of the radiation is affected by the location of the ZNP. When the slope of the nonlinear parameter $\gamma$ is small, the location of the radiation frequency $\Omega_{DW}$ hardly depends on the ZNP. However, there is a radical change in the radiation frequency when the ZNP is kept close to the radiation frequency. Numerically we observe that the ZNP acts as a strong barrier when Raman term is present in the GLE, and the radiation frequency can never cross the ZNP. However, in absence of Raman term (i.e. when only TOD term is present) the radiation can crosses the ZNP. If we gradually change the value of $\mu_{1eff}$ such that the ZNP ($\Omega_{ZNP}$) shifts towards the radiation frequency, then the radiation frequency is also shifted towards the input frequency. It seems like the ZNP drags the radiation towards the input frequency. A sudden change of radiation frequency is observed when the ZNP crosses the input frequency. That means now the ZNP and zero GVD point are in opposite side of the input frequency. Note that, for such condition the DS is excited in same regime ($\beta_2>0$ and $ \gamma_{0eff}<0$), but slope of the nonlinear dispersion profile is changed. The change of radiation frequency $\Omega_{DW}$ with respect to ZNP is shown in Fig.\,\ref{variational_all_TOD}(c). The solid line corresponds to the results that we obtain from our derived PM condition, whereas the stars represent the numerical data. Further we perform a simulation where we gradually shift the zero GVD point towards negative frequency keeping the ZNP fixed. We note that the ZNP reduces the radiation frequency compared to its actual location. Here the \textit{actual} location means the frequency location of the radiation in the absence of ZNP. In order to visualize this effect we plot $\Omega_{DW}$ as a function of $\delta_3$ in Fig.\,\ref{variational_all_TOD}(d) in presence (red stars) and in absence (blue dots) of ZNP. It is evident that the radiation frequency is reduced in the presence of ZNP, which is in good agreement with the PM equation derived by us (solid red line).

\section{Conclusions}
In this brief report, we demonstrate that the formation of DS is versatile and it can be excited in self-focusing and self-defocusing nonlinear systems with ND and AD. It is noted that the excited DSs exhibit identical nature in the domain where the numeric sign of the multiplication of nonlinear coefficient ($\gamma_{0eff}$) and GVD coefficient ($\beta_2$) is same. We numerically find when DSs are excited in the regime $\gamma_{0eff}\beta_2>0$, the pulse  is quite robust with an unusual flat-top spectra. We further investigate the influence of the higher-order perturbations on the dynamics of DS in the presence of ZNP. We adopt variational treatment to understand the dynamics of the individual pulse parameter during propagation. An extensive calculation leads to a set of coupled differential equation, which are essentially the equation of motion of the pulse parameters. The influence of ZNP is found to be significant, as it behaves like a barrier that does not allow any radiation to pass through it in the presence of IRS. The Raman redshift is arrested by the ZNP and saturates to a given frequency $\Omega_{p\,sat}$. This happens when the ZNP and Raman red-shifted frequency are in the same side of the input frequency. The situation is reversed when the Raman red-shifted frequency and the ZNP are in the opposite sides of the input frequency. In such a situation the ZNP pushes the Raman frequency more towards lower frequency side, and we obtain enhanced red-shifting.  The variational treatment theoretically predicts the $\Omega_{p\,sat}$ that matches well with the numerical data. A similar phenomenon is observed for TOD mediated radiation under ZNP. The location of the radiation frequency is influenced by the ZNP. We establish an analytic expression that predicts the radiation frequency in the presence of ZNP. It is considered that the phase-matched radiation is sandwiched between ZNP and zero GVD point. When the ZNP is moved towards zero GVD point, it (ZNP) drags the radiation towards zero GVD point. The concept of negative nonlinearity is relatively new for DS, and it produces some interesting effects like formation of DS in ND, and modulation of Raman redshift and phase-matched radiation through ZNP. In our work, we aim to explore the behaviour of DS under self-defocusing nonlinearity and the role of ZNP.  The existence of the negative nonlinearity adds new degrees of freedom in manipulating the pulse dynamics in waveguides and can be useful in practical applications.

\vspace{-0.2cm}
\section*{ACKNOWLEDGMENTS}       
 
A.S. acknowledges MHRD, India for a research fellowship.


\end{document}